\documentclass[twocolumn,showpacs,preprintnumbers,amsmath,amssymb,prl]{revtex4}
\usepackage{graphicx}
\usepackage{dcolumn}
\usepackage{bm}
\begin{document}
\title{Strong correlation effects and optical conductivity in
electron doped cuprates}
\author{Tanmoy Das, R. S. Markiewicz, and A. Bansil}
\address{Physics Department, Northeastern University, Boston MA
02115, USA}
\date{\today}
\begin{abstract}

We demonstrate that most features ascribed to strong correlation effects
in various spectroscopies of the cuprates are captured by a calculation of
the self-energy incorporating effects of spin and charge fluctuations. The
self energy is calculated over the full doping range of electron-doped
cuprates from half filling to the overdoped system. The spectral function
reveals four subbands, two widely split incoherent bands representing the
remnant of the split Hubbard bands, and two additional coherent, spin- and
charge-dressed in-gap bands split by a spin-density wave, which collapses
in the overdoped regime. The incoherent features persist to high doping,
producing a remnant Mott gap in the optical spectra, while transitions
between the in-gap states lead to pseudogap features in the mid-infrared.

\end{abstract}
 \maketitle\narrowtext

A key issue in cuprate physics is to understand the routes through which
cuprates evolve from a `Mott' insulator at half filling to a high-$T_c$
superconductor at optimal doping.  Underdoped cuprates show the presence
of a `pseudogap' or a region of depleted density of states extending to
several hundred meV's around the Fermi energy and thus are very
unconventional, but the overdoped system behaves closer to a normal Fermi
liquid and is thus amenable to a largely conventional description. Along
these lines, evidence of small Fermi surface pockets\cite{leyraud} has
been reported in the underdoped regime, while in overdoped
Tl$_2$Ba$_2$CuO$_{6+\delta}$, a large, three-dimensional Fermi surface
consistent with LDA band structure calculations\cite{hussey} is observed.
In sharp contrast, optical studies point to a far more complex and
puzzling picture in that even in the overdoped case, an absorption peak
characteristic of the Mott gap continues to persist\cite{onoseprb,uchida}
in the spectra, suggesting that doping introduces new in-gap states in
which the pseudogap physics resides, but that otherwise the Mott gap
persists at all dopings.  Here we show that a relatively simple and
transparent model of the electronic self energy, where the quasiparticles
are dressed with spin wave excitations, captures the key experimentally
observed features of the remarkable doping evolution in optical and other
spectroscopies, including in particular the persistence of the Mott gap in
the overdoped regime. Our self-energy is also in good accord with quantum
Monte Carlo computations in the parameter regimes where the latter are
available.

We evaluate the self-energy $\Sigma$ as a convolution over
the green function $G$ and the interaction $W\sim U^2\chi$ (including
both spin and charge contributions) as \cite{vignale},
\begin{eqnarray}\label{selfeng}
\Sigma(\vec{k},\sigma,i\omega_n)=\frac{3}{2}U^2
\sum_{\vec{q},\sigma^{\prime}}^ {\prime}
\int_{-\infty}^{\infty}\frac{d\omega_p}{2\pi}
~~~~~~~~~~~~~~~~~~~~~\nonumber\\
G(\vec{k}+\vec{q},\sigma^{\prime},\omega_n+\omega_p)
\Gamma(\vec{k},\vec{q},\omega_n,\omega_p){\rm
Im}[\chi^{\sigma\sigma^{\prime}}(\vec{q},\omega_p)],
\end{eqnarray}
where $\sigma$ is the spin index and the prime over the
$\vec{q}$ summation means that the summation is restricted to the
magnetic Brillouin zone. In the underdoped region, the pseudogap
is modeled by an antiferromagnetic
(AFM) order parameter, resulting in $G$, $\chi$ and $\Sigma$
becoming $2\times2$ tensors\cite{SWZ}. We define a total
self-energy as $\Sigma^{t}=US\tilde{\tau_1}+\Sigma$, where
$\tilde{\tau_1}$ is the Pauli matrix along the $x-$direction and
$US$ is the AFM gap defined below. The self-energy $\Sigma^t$
contains essentially two energy scales: (i) it gives rise to the SDW
with an additional renormalization of the overall quasiparticle
dispersions in the low energy region, and (ii) at higher energies it
produces the Hubbard bands.  We use a modified
self-consistent scheme, referred to as quasiparticle$-GW$
(QP$-GW$)-scheme in which
$G$ and $W$ are calculated from an approximate self-energy
$\Sigma^{t}_0(\omega)=US\tilde{\tau_1}+
\left(1-Z^{-1}\right)\omega\tilde{\bf{1}}$, where the
renormalization factor $Z$ is adjusted self-consistently to match
the self-energy $\Sigma^{t}$ at low
energy.\cite{markiecharge,markiewater}

We find that near optimal doping spin
waves\cite{markiewater,markiecharge,macridin,moritz,foot0} dress the
quasiparticles into a coherent {\it in-gap state}, while the incoherent
high-energy features are remnants of the upper and lower Hubbard bands
(U/LHBs)\cite{foot1}. With underdoping the in-gap state develops into a
spin density wave (SDW) state which opens a gap between the upper and
lower magnetic bands (U/LMBs). The model also describes the high-energy
kink or the waterfall effect seen in the electronic
dispersion\cite{ronning} as the crossover between coherent and
incoherent features.

The present calculations are restricted to the electron-doped cuprates in
order to avoid possible complications of nanoscale phase separation.
The self-energy $\Sigma_0^t$ splits the LDA-band, $\xi_{\vec{k}}$
(modelled by tight-binding (TB) parameters\cite{markiewater}) into
renormalized UMB ($\nu=+$) and LMB ($\nu=-$):
\begin{equation}\label{band}
E_{\vec{k}}^{\nu}=Z\left(\xi_{\vec{k}}^{+}\pm\sqrt{(\xi_{\vec{k}}
^{-})^2+(US)^2}\right),
\end{equation}
where $\xi_{\vec{k}}^{\pm}=(\xi_{\vec{k}}\pm\xi_{\vec{k}+\vec{Q}})
/2$. The AFM magnetization $S$ at $\vec{Q}=(\pi,\pi)$ is calculated
self-consistently at each doping, assuming a doping dependent
$U$.\cite{kusko}  In the present formalism $U$ is renormalized by
$Z$. The doping dependency of $U$ is chosen such that $ZU$
reproduces the pseudogap in both angle-resolved photoemission
spectroscopy (ARPES) and optical spectra, while $x$, $S$, and $Z$ are
determined self-consistently.\cite{tanmoy2gap} Finally, the vertex correction
$\Gamma(\vec{k},\vec{q},\omega,\omega_p)$ in Eq. \ref{selfeng} is
taken as its first order approximation (Ward's identity) as
$\Gamma(\vec{k},\vec{q},\omega,\omega_p)=1/Z$. Since the
$k$-dependence of $\Sigma$ is weak, we further simplify the
calculation by assuming a $k$-independent $\Sigma$, which we
calculate at a representative point $k=(\pi/2,\pi/2)$.
\begin{figure}
\rotatebox{0}{\scalebox{0.45}{\includegraphics{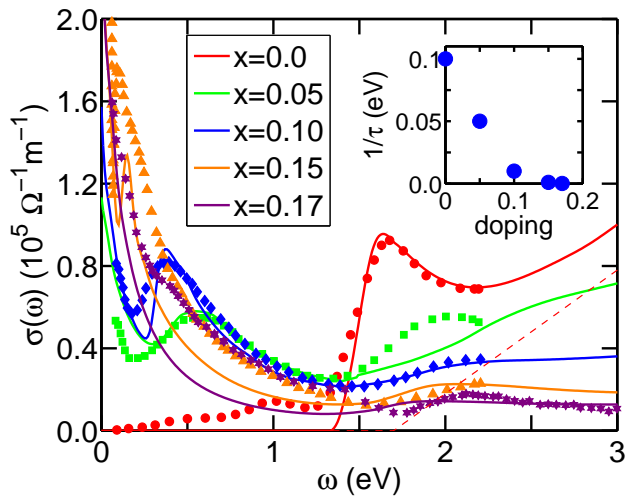}}}
\caption{(color online) Calculated optical conductivity (solid lines)
compared
with experiment (symbols of same color) for dopings
$x=0.0$ to $x=0.15$ as taken from Ref.~\onlinecite{onoseprb}
and for $x=0.17$ from Ref.~\onlinecite{uchida}. Experimental
data for $x=0.17$ is subtracted by a background contribution
to match the other data set.
 The dashed line gives the background
contribution added to the theoretical spectrum at $x=0$.
 {\it inset:} Inferred doping dependence of the scattering rate.}
\label{se}
\end{figure}

We begin by discussing the optical spectra of Fig.~1.  The frequency
dependent optical conductivity, $\sigma (\omega)$, is calculated using
standard linear response theory in the AFM state,\cite{allen} for the full doping
range from the half-filled state to the quantum critical point (QCP) in the
overdoped region.  A very good level of accord is seen with the
experimental results\cite{onoseprb,uchida}. To fit the Drude conductivity, we
have introduced an impurity scattering rate $\tau$ which is found to have a strong doping
dependence (inset). At high energy we include a doping-dependent background
contribution, presumably associated with interband transitions to
higher-lying bands not included in the present calculations. We have
used the same energy-dependence of the background for all dopings
with an intensity that decreases smoothly with doping.  The red
dashed line in Fig.~1 shows this interband contribution for the
$x=0.0$ spectrum.

Interestingly, the spectra show a nearly isosbetic (equal absorption)
point near 1.4 eV, consistent with the experimental behavior. The
doping evolution is completely different on opposite sides of this isosbetic
point. Above this point the spectrum is dominated by a broad hump feature
above 1.5 eV, a signature of the Mott gap.  At half-filling, only this
feature is present and the optical spectrum shows an insulating gap whose energy, structure,
and intensity match remarkably with measurements\cite{onoseprb}.  As
doping increases the high energy peak shifts to higher energy and
broadens.

Below the isosbectic point there is little spectral weight at half
filling, but as doping increases spectral weight is gradually
transferred from the higher energy region to the mid-infrared (MIR) one.
The lower energy spectrum is associated with a Drude peak related to
intraband transitions and a mid-infrared peak associated with transitions across the magnetic
gap [the pseudogap for the electron-doped cuprates].  With doping,
this peak shifts to lower energy as the magnetic gap collapses
and gradually sharpens due to the doping dependent
scattering rate, see inset to Fig.~1. Note that at $x=0.17$, when
the pseudogap has collapsed, Mott-gap features still persist in the
spectrum.  The present mean-field calculation overestimates the
N\'eel temperature $T_N$, but this can be corrected by including
critical fluctuations, {\it grave a la} the Mermin-Wagner
theorem\cite{markieMW}.

\begin{figure}
\rotatebox{0}{\scalebox{0.4}{\includegraphics{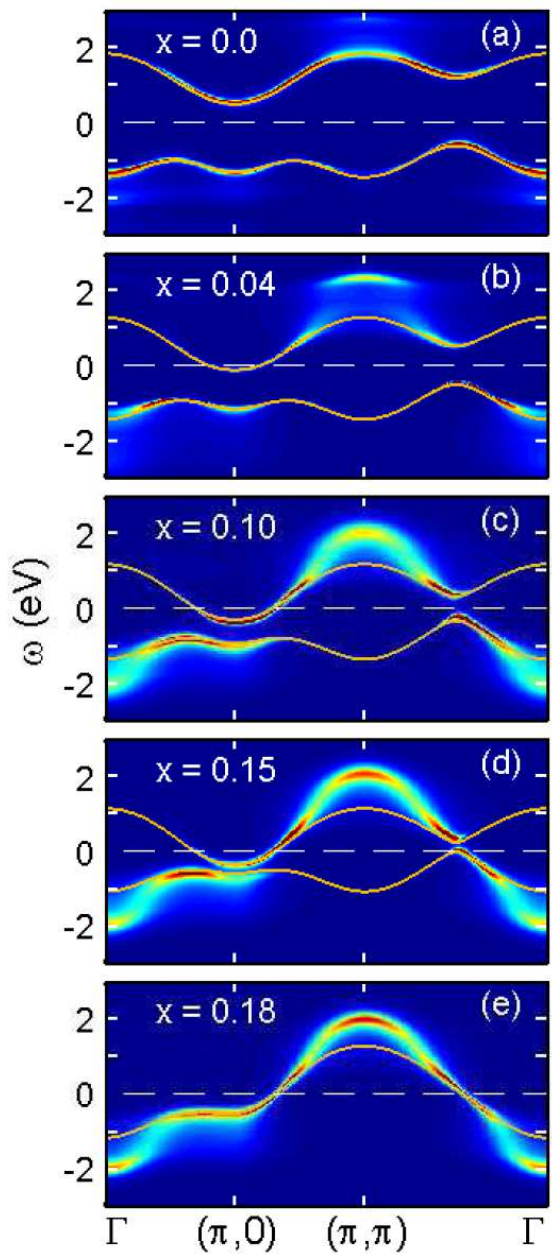}}}
\caption{(color online) Spectral intensity as a function of $\omega$
along the high-symmetry lines for several dopings (at temperature
$T=0$). Blue to red color map gives
the minimum to maximum intensity.
 In each panel, the gold lines represent the renormalized
magnetic bands ($\Sigma_0$-dressed).} \label{spectral}
\end{figure}

The origin of these features can be understood by looking at the doping
dependence of the momentum-resolved spectral weight in Fig.~2.  In the
overdoped case in Fig.~2(e), a kink due to the bosonic coupling reproduces
the waterfall effect below $E_F$, with a corresponding effect above $E_F$,
splitting the spectrum into an effective three-band behavior, with UHB,
LHB, and in-gap states.  The features in the optical spectra are
associated with transitions between these bands: The residual Mott gap
arises from the transition from the LMB to the incoherent UHB [or from LHB
to UMB], while the Drude term is associated with intraband transitions near
the Fermi level.  At lower doping an AFM gap opens in the coherent in-gap
states\cite{foot3}, leading to the UMB/LMB splitting and a four band
behavior similar to that seen in QMC cluster calculations\cite{grober}.
Consistent with the QMC calculations, the coherent in-gap bands are
dressed by magnetic quasiparticles.  As the magnetic gap opens, the MIR
feature in the optical spectra, being associated with transitions across
this gap, shifts to higher energy.

The quality of self-consistency of our scheme can be assessed by noting
that the final coherent bands have nearly the same dispersion as the
$\Sigma_0-$dressed ones (gold lines) used as input to obtain the
self-energy. The doping dependence of the two coherent magnetic
bands is in excellent agreement with experiments\cite{nparm} and
earlier mean field calculations\cite{kusko}, and captures the
incoherent weight (the UHB/LHBs) at higher energies seen
experimentally. Note that the incoherent weight is concentrated near
the top and bottom of the bare LDA bands, leading to a nearly doping
independent UHB-LHB splitting.
\begin{figure}
\rotatebox{270}{\scalebox{0.42}{\includegraphics{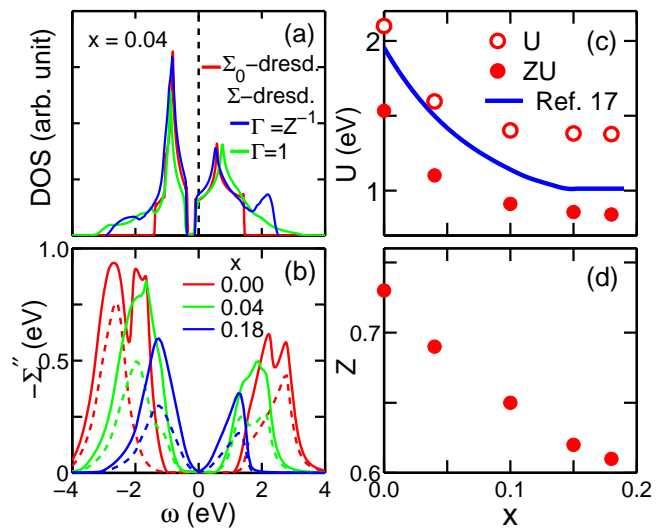}}}
\caption{(color online) (a) The QP-GW DOS (blue lines) is compared
with $\Sigma_0$-dressed DOS (red lines) calculated at $T=0$. The green lines show the DOS
without the vertex correction. (b) Solid lines give the
imaginary part of the total self-energy, while the dashed lines
give the corresponding charge contributions. (c) Doping dependence
of
self-consistent values of $U$ and $ZU$ is compared with earlier mean
field results\cite{kusko,tanmoyprl}. (d) Renormalization factor $Z$
decreases linearly with doping.} \label{dos}
\end{figure}

Further insight is provided by Fig.~3(a), which compares the full
QP-GW DOS (blue line) and the $\Sigma_0$-dressed quasiparticle DOS
(red line), normalized to the same peak height, for a representative
doping $x=0.04$.  The good agreement between various computations over
most of the energy range indicates the
high degree of self-consistency in the self-energy. The dressed DOS shows
four well-separated peaks. A clear leading edge gap of $\sim$0.3~eV
($\sim$1~eV at $x=0$) can be seen. A gap persists at higher doping
up to $x=0.18$ (Fig.~1), although it is obscured in the DOS by band overlap. The
importance of the vertex correction is illustrated by the green line
in Fig.~3(a), which shows that setting $\Gamma=1$ reduces the weight
in the U/LHBs.

Figure~3(b) shows how the calculated imaginary self-energy
$\Sigma^{\prime\prime}$ evolves with doping. The solid lines give
the total $\Sigma^{\prime\prime}$, with the corresponding dashed
lines giving the charge contribution.
[$\Sigma_{\rm{spin}}=\Sigma_{\rm{total}}-\Sigma_{\rm{charge}}$ is not
shown]. In the underdoped region the extra splitting at high energies in
the self energy is related to spin-charge separation. The
spin response is significant at low energy for all doping but the
charge contribution is nearly zero in the low energy region for the
lower doping and becomes finite only at higher energies above $\sim
3$~eV. As doping increases, the charge response moves toward the
Fermi level and increases in contribution to its total value.
At $x=18\%$, when the AFM gap
vanishes, the charge and spin susceptibility become equal.  Note that
the broadening of the self-energy evident in Fig.~3(b) is reflected in the
increasing broadening of the Mott gap feature with doping. The shift of
the peak in the imaginary part of the
self-energy towards $E_F$ in Fig.~3(b) reflects the doping dependence
of the MIR feature in Fig.~1.

Fig.~3(c) describes the doping dependence of $U$. Although the
renormalized Hubbard parameter $ZU$ follows almost the same doping
dependence as in mean field calculations\cite{kusko,tanmoyprl}, the
bare $U$ displays considerably weaker doping dependence away from
half-filling.  The renormalization factor $Z$ in Fig.~3(d) actually
increases
with underdoping.\cite{foot4}  This can be readily
understood. As we lower doping, the AFM gap increases leading to
a decrease of the spectral weight near the Fermi level. This causes
a reduction of the real part of the self-energy, shifting the peak
in $\Sigma^{\prime}$ towards higher energy.  The resulting slope
decrease leads to a larger $Z$.
\begin{figure}
\rotatebox{270}{\scalebox{0.45}{\includegraphics{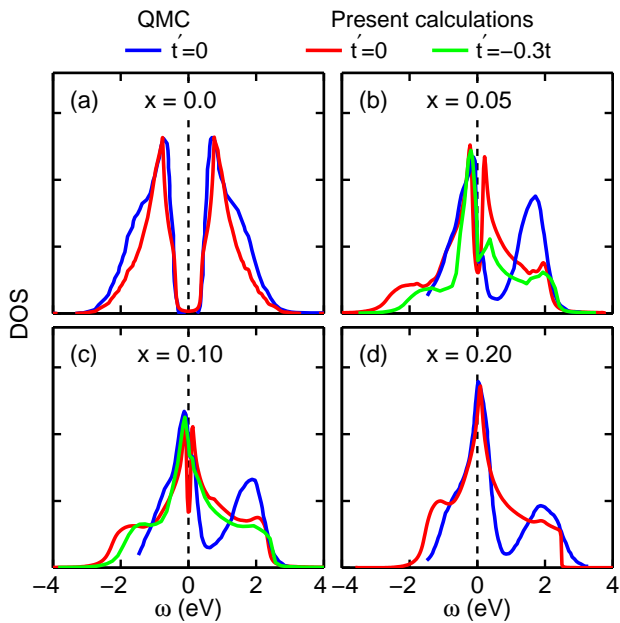}}}
\caption{(color online) The computed DOS at various dopings are
compared with the corresponding QMC results (blue lines) for
$x=0.0$\cite{maier} and for $x=0.05$ to $x=0.20$ \cite{jarrell}.
The red
lines in each panel give our result for $t^{\prime}=0$, whereas the
green lines in (b) and (c) are for $t^{\prime}=-0.3t$.} \label{qmc}
\end{figure}

Existing QMC calculations on the cuprates generally employ simpler
hopping parameter sets, so we have repeated our calculations with
the same band parameters for a quantitative comparison in Fig.~4. All
the QMC results (blue lines) are obtained for $t^{\prime}=0$,
$U=8t$ with a momentum dependent self-energy correction\cite{maier,jarrell}.
We have obtained the corresponding DOSs (red lines) for $t^{\prime}=0$
 for $\Sigma$ calculated at a fixed momentum of $k=(\pi/2,\pi/2)$,
using a doping dependent renormalized $U$ ($ZU$)
very similar to the one found for the electron doped case
in Fig.~3(c). [Note that the QMC automatically generates a
renormalized $U$]. Our result reproduces the QMC very well for $x=0$
in Fig.~4(a) where a prominent electron-hole symmetry is observed. At
higher doping, the QMC results show a relatively smaller coherent
peak above the Fermi level, whereas our result continues to exhibit
near electron-hole symmetry in the in-gap region as might be
expected for $t^{\prime}=0$ dispersion.  We can mimic a momentum
dependent self-energy by including $t^{\prime}=-0.3t$, and find
that this provides better agreement with the QMC results as shown
by the green lines in Figs.~4(b) and 4(c).  Finally,
at $x=0.20$, the pseudogap collapses and our result with
$t^{\prime}=0$ agrees very well with the QMC in Fig.~4(d).  As
might be expected from our approximate self-energy calculation, the
weight of the UHB is generally underestimated.

In summary, we find that spin-wave dressing of the quasiparticles explains
the incoherent U/LHB features seen in various experiments including the
recent experiments observing waterfall effects in the cuprate spectra. The
self-energy corrections not only renormalize the large widths of the LDA
dispersions but also restore the residual incoherent spectral weight
associated with U/LHBs. In the underdoped regime, we show that the
coherent in-gap bands reproduce both the four-band behavior seen in
quantum cluster calculations and the magnetic gap collapse found in the
mean-field calculations and a variety of experiments. The puzzling
persistence of the Mott gap in optical spectra, even as the magnetic gap
collapses, is thus reconciled. The fact that our calculations work so well
confirms that the cuprates can be understood within the intermediate coupling regime,
with $U$ much less than twice the bandwidth.\cite{comanac}.

\begin{acknowledgments}

We thank Mark Jarrell for important conversations. This work is supported
by the U.S.D.O.E, Basic Energy Sciences, contracts DE-FG02-07ER46352 and
DE-AC03-76SF00098, and benefited from the allocation of supercomputer time
at NERSC and Northeastern University's Advanced Scientific Computation
Center (ASCC).

\end{acknowledgments}


\begin{thebibliography}{99}
\bibitem{foot0} While the low-energy carriers are mainly dressed by
electronic excitations in the magnetic
channel\cite{markiewater,macridin},
the charge channel can also contribute\cite{markiecharge}.
\bibitem{foot1} In reality, the Mott physics lies at a higher energy,
comparable to the bonding-antibonding separation of the CuO$_2$-hybridized
bands, and the bands we refer to as U/LHB are the incoherent high
energy tails of the antibonding (LDA) band.
\bibitem{foot3} We use a single renormalisation factor for both
bands at all dopings.
\bibitem{foot4}Recall that this $Z$ parameter represents an average
renormalization over the coherent band, and is not necessarily equal
to the conventional renormalization factor at $E_F$.
\end{thebibliography}
\end{document}